\documentclass[]{spie}

\usepackage{amsmath,amssymb,graphicx}
\usepackage[colorlinks=true,allcolors=blue]{hyperref}

\newcommand{\ed}{\mathop{}\!\mathrm{d}}

\title{The Black Hole Explorer: Using the Photon Ring to Visualize Spacetime Around the Black Hole}

\author[a]{Peter Galison}
\author[a,b]{Michael D. Johnson}
\author[c]{Alexandru Lupsasca}
\author[c]{\\Trevor Gravely}
\author[c]{Roman Berens}
\affil[a]{\small Black Hole Initiative at Harvard University, 20 Garden Street, Cambridge, Mass., USA}
\affil[b]{Center for Astrophysics $\vert$ Harvard and Smithsonian, 60 Garden Street, Cambridge, Mass., USA}
\affil[c]{Department of Physics \& Astronomy, Vanderbilt University, Nashville, Tenn., USA}

\authorinfo{Send correspondence to: galison@fas.harvard.edu, mjohnson@cfa.harvard.edu, alexandru.v.lupsasca@vanderbilt.edu, trevor.gravely@vanderbilt.edu, and roman.berens@vanderbilt.edu}

\begin{document} 
\maketitle

\begin{abstract}
The Black Hole Explorer (BHEX), an orbiting, multi-band, millimeter radio-telescope, in hybrid combination with millimeter terrestrial radio-telescopes, is designed to discover and measure the thin photon ring around the supermassive black holes M87* and Sgr\,A*.
As background to the BHEX instruments, this paper explores various aspects of the photon ring, focusing on the intricate flow of light around a spinning black hole, and tracking, through visual simulations, photons as they course along geodesics.
Ultimately, the aim of these visualizations is to advance the foundational aims of the BHEX instrument, and through this experiment to articulate spacetime geometry via the photon ring.
\end{abstract}

% Include a list of keywords after the abstract 

\section{Introduction}

Orbiting at an altitude of 20,000km, the Black Hole Explorer (BHEX) will carry a multi-band, millimeter radio-telescope, forming a hybrid observatory with terrestrial radio-telescopes (including the Event Horizon Telescope and its future development).
In the first instance, BHEX is designed to discover and measure the thin photon ring around the supermassive black holes M87* and Sgr\,A*.\cite{Johnson2020,Johnson2024,Lupsasca2024}

\begin{figure}[h]
	\centering
	\includegraphics[width=.6\textwidth]{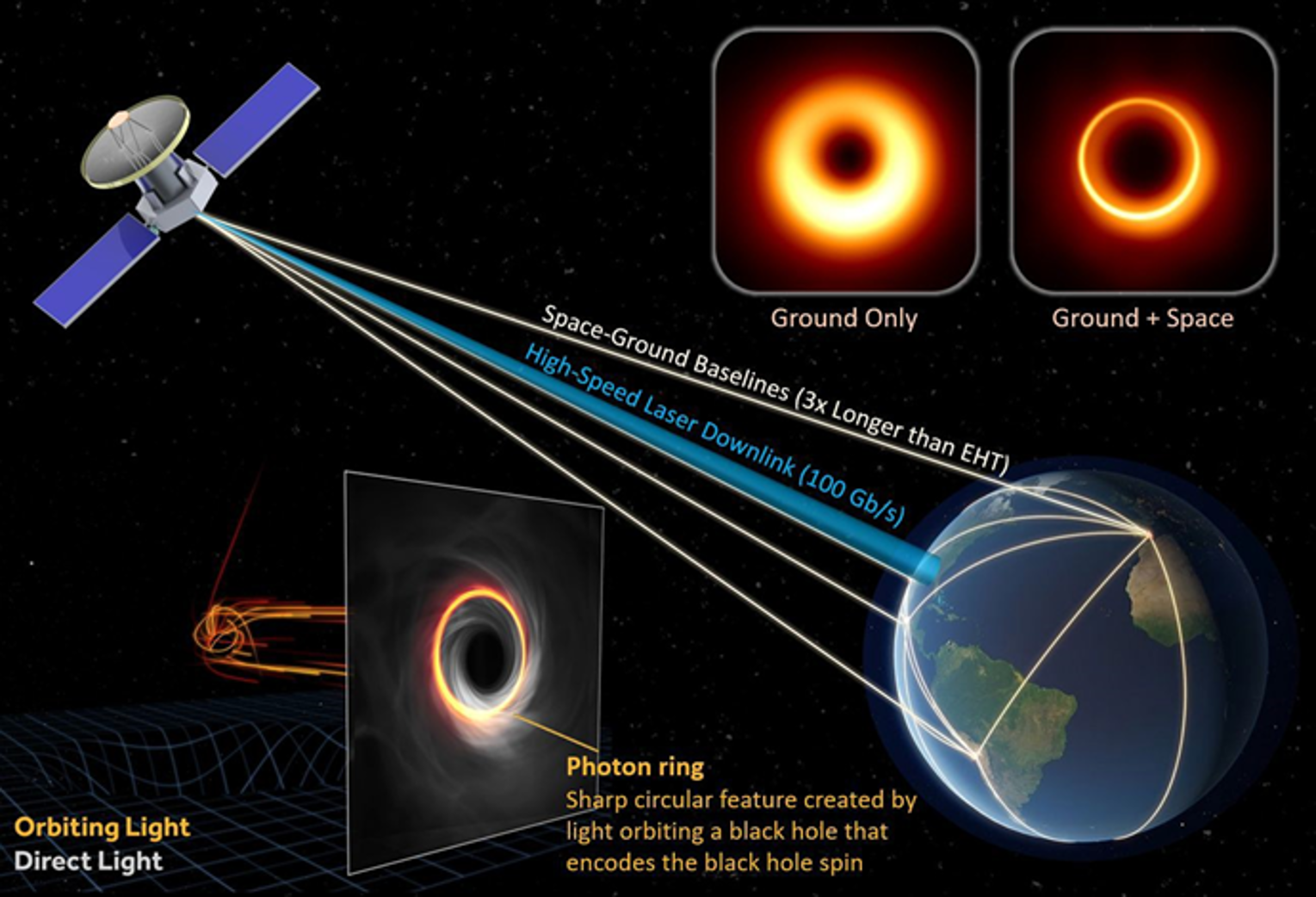}
	\caption{The BHEX mission concept. Reproduced from [\citenum{Johnson2024}].}
	\label{fig:Concept}
\end{figure}

\begin{figure}[t]
    \centering
    \includegraphics[width=\textwidth]{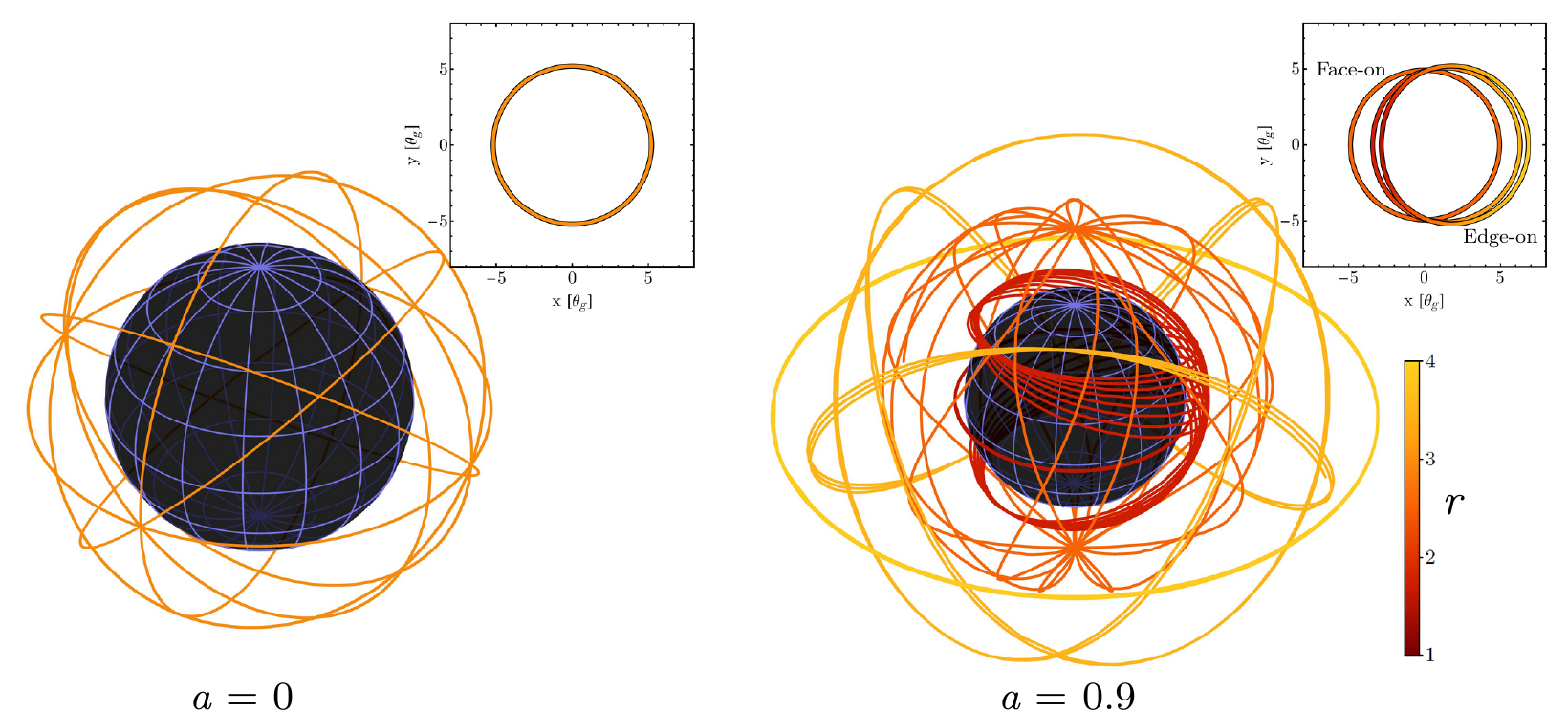}
    \vspace{10pt}
    \caption{
    {\bf The photon ring.} The left panel shows photon orbits around a Schwarzschild black hole, which must lie at an orbital radius $r\equiv3GM/c^2$ (defining the photon sphere).
    The image of asymptotically bound photons gives rise to a circular photon ring (shown as the inset).
    The right panel shows photon orbits for a rapidly rotating black hole (with spin $a/M=90\%$), which fill in a volume of spacetime known as the photon shell.
    The corresponding photon ring is both smaller and deformed compared to that of the Schwarzschild black hole.
    Reproduced from [\citenum{Johnson2024}].
    }
\label{fig:PhotonRing}
\end{figure}

Formed by light lensed and thrown into near-horizon orbit by gravity, these orbiting photons carry information about the black hole (including mass and spin).
The region around the black hole where these light-orbits occur is known as the photon sphere---a special region of spacetime close to the horizon.
Photons in such orbits within this photon sphere are perched on a knife's edge between falling into the black hole or flying off to infinity.
For a non-spinning (Schwarzschild) black hole, the radius is fixed---we would observe a perfect ring.
For a spinning (Kerr) black hole, the thickness of the photon sphere (and the radius of its inner edge) is variable, defining the photon shell; on an image, we would observe an asymmetric ring of inhomogeneous shape, thickness and brightness that depends both on inclination of viewing relative to the black hole spin, and the spin magnitude of the black hole.
Visualizations of these spacetime quantities and their corresponding images are shown in Fig.~\ref{fig:PhotonRing}.

Only rarely do we have the opportunity to simultaneously probe both fundamental physics and astrophysics.
The photon ring offers just such an opportunity.
On the astrophysics side, spin offers our best candidate as the underlying dynamo for launching galaxy-spanning jets that can emerge from black holes like M87*.
And yet, even though spin is one of the fundamental characteristics of a black hole, measuring this quantity has proven difficult because of its dependence on detailed modeling of accretion dynamics; not so for the photon ring.

At the same time, the photon ring offers a glimpse into fundamental physics.
Photons flow along null geodesics; the photon ring reveals the geometry of spacetime bordering the black hole horizon---only very weakly affected by the the accretion disk.\cite{Johnson2020}
One of the most striking features of the photon sphere of a Kerr black hole is that the spacetime itself is prone to chaos, in the sense that null geodesics initially close to one another separate, radically, developing into wildly divergent orbits.
(The quantity characterizing this chaotic behavior is the Lyapunov exponent, named for Alexandr Lyapunov, who introduced it to characterize unstable fluid flow.)

BHEX will give us our first measurements of spacetime chaos near a black hole.
This paper, together with its time-dependent simulations of initially nearby photons within the photon ring, offers a visual grasp of the chaotic spacetime near the horizon, a Kerr black hole, an edge of the universe.   

\section{``Adjacent'' paths diverge (by radius)}

\begin{figure}[h]
	\centering
	\includegraphics[width=.5\textwidth]{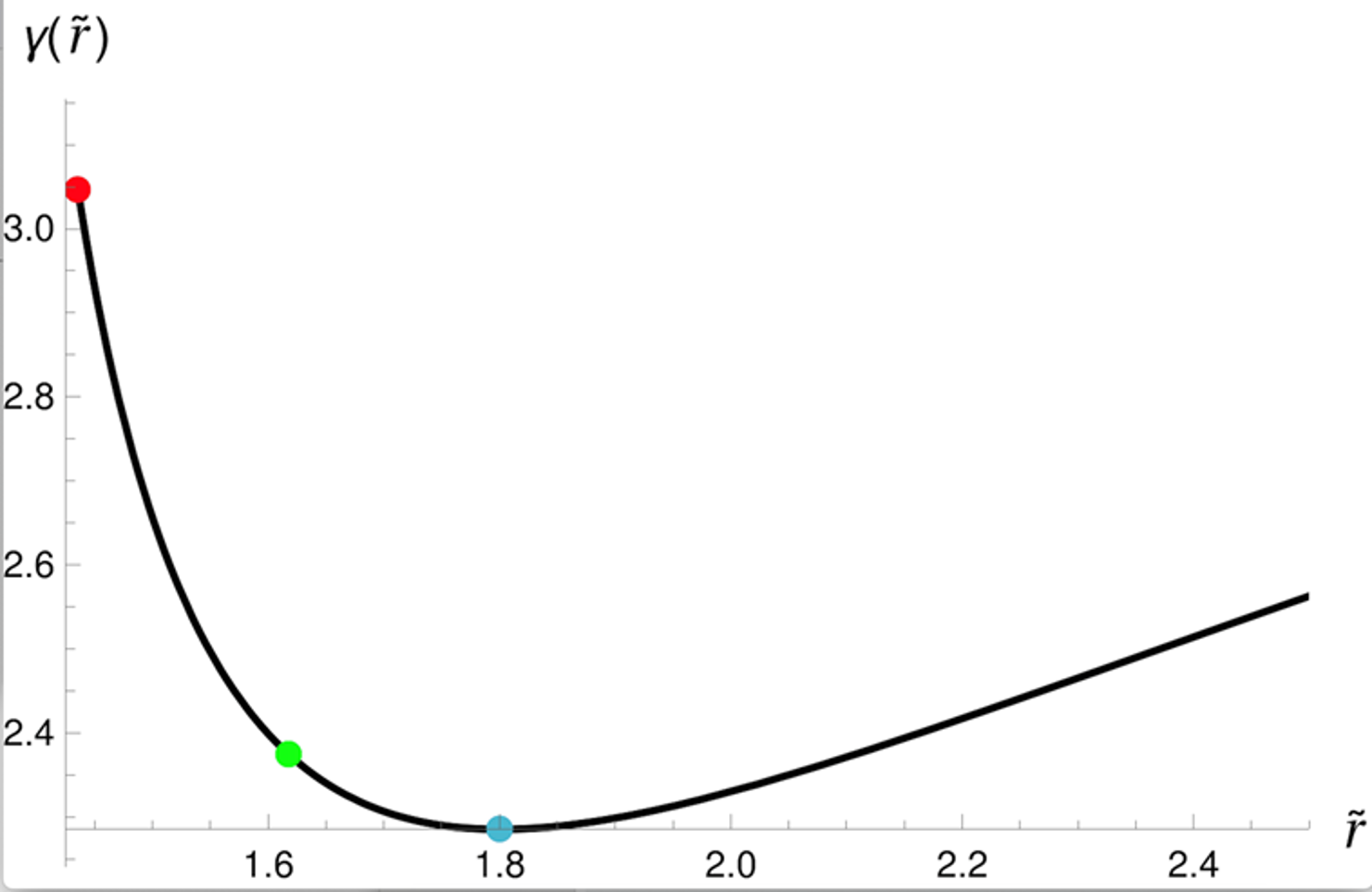}
	\caption{The Lyapunov exponent, $\gamma$, characterizes the onset of chaos through the divergence of nearby trajectories in phase space.
	This figure shows $\gamma$ as a function of orbtial radius $\tilde{r}$ for a rapidly spinning black hole, from the inner to the outer edge of the photon shell---at these edges, there is more chaotic behavior.
	Between these two edges lies a minimum, where the rate of separation (chaoticity) is minimized.
	The dot colors correspond to the trajectory colors in Fig.~\ref{fig:Adjacent}.}
	\label{fig:Exponent}
\end{figure}

An infinite series of self-similar rings can be indexed by $n$, the number of half-orbits that photons make around the black hole before escaping to us.
From the ratio of the diameter of the $n=1$ ring (composed of photons that make one half of a turn before escaping to our instruments) to the diameter of the $n=0$ ring (photons that travel to us directly), we can deduce the Lyapunov exponent, $\gamma$ (see Fig.~\ref{fig:Exponent}).
Or, with greater precision, $\gamma$ can be obtained from the ratio of the $n=2$ ring diameter to the $n=1$ ring diameter.

While BHEX has distinct astrophysical goals (including the first direct measurement of black hole spin, the probing of jet formation, and the study of numerous active galactic nuclei), the mission will begin a probe of near-horizon spacetime itself through the photon ring, which arises from unstable spherical photon orbits.\cite{Johnson2020}
Although geodesic motion in the exact spacetime of a Kerr black hole is integrable, we expect the (nearly) bound photon orbits to become chaotic under small perturbations of the background.
Here, we visualize the tomography of this incipient chaos by simulating the separation of nearby photon trajectories and studying their rate of separation.

\begin{figure}[h]
	\centering
	\includegraphics[width=.7\textwidth]{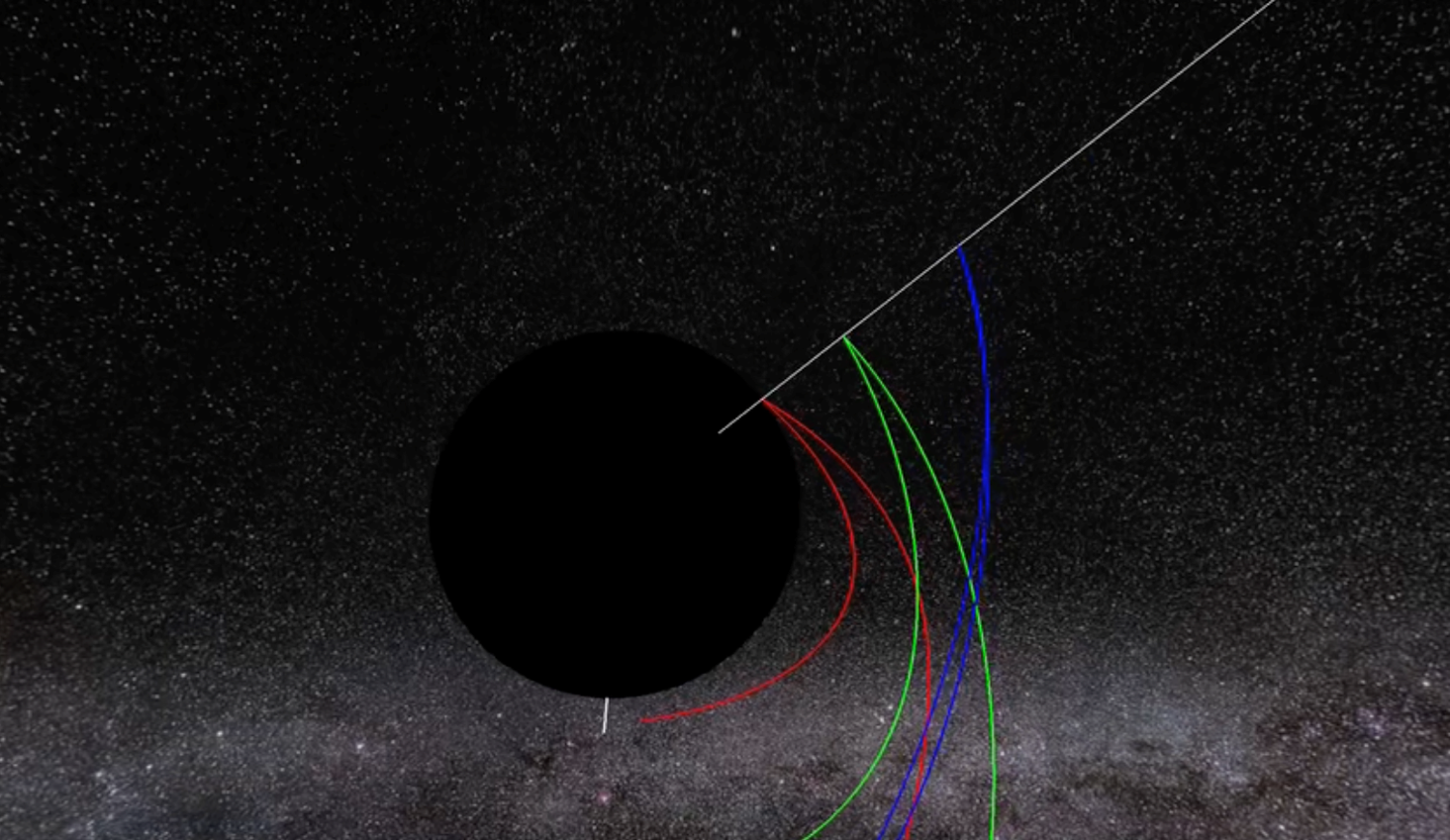}
	\caption{``Adjacent'' photon trajectories near a rapidly spinning black hole (with spin $a/M=0.94\%$) at three different radii from the horizon.
	The colors match those of the points on the diagram in Fig.~\ref{fig:Exponent}---blue is the chaotic minimum.}
	\label{fig:Adjacent}
\end{figure}

\section{Subcritical, critical, supercritical geodesics}

\begin{figure}[h]
	\centering
	\includegraphics[width=.8\textwidth]{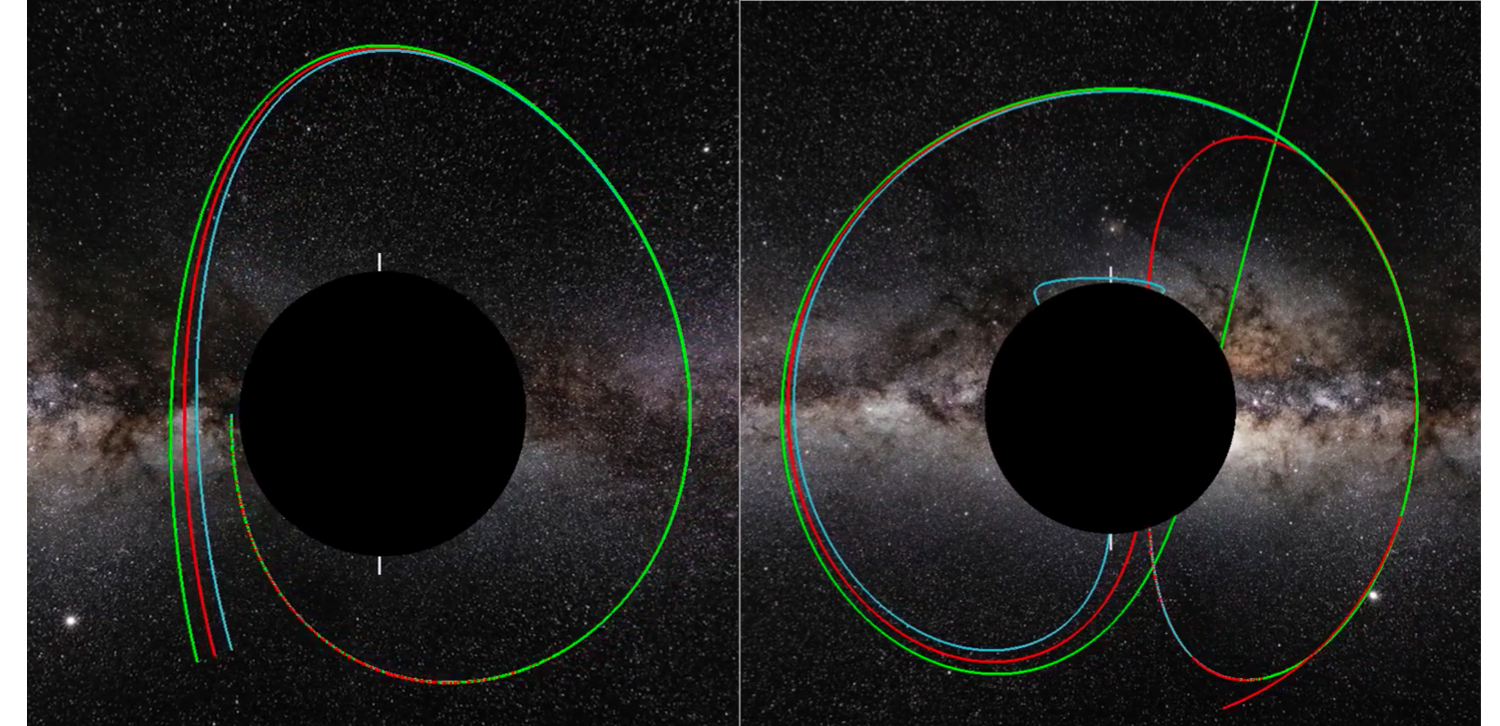}
	\caption{In these two frames from an animated simulation, three nearby photons, separated by small differences in radius $r$, initially head off on almost indistinguishable trajectories.
	The red photon is critical (remains in orbit), the subcritical blue photon eventually falls into the black hole, and the supercritical green photon escapes to infinity.
	This is a demonstration of the butterfly effect in spacetime itself: small initial differences in geodesics lead to noticeable differences within a single orbital period (left frame) and dramatically different fates at later times (right frame).}
	\label{fig:Differences}
\end{figure}

The photon ring asymptotes to a ``critical curve'': the infinitely thin image of trajectories along which photons eventually enter---and thereafter remain in---orbit. (Here the black hole spin is $a/M=0.94\%$.)
Inside the critical curve, photons fall into the black hole; outside of it, they escape to infinity (see Fig.~\ref{fig:Differences}).

\section{``Real-space'' diagram vs. embedding diagram}

\begin{figure}[!ht]
	\centering
	\includegraphics[width=.75\textwidth]{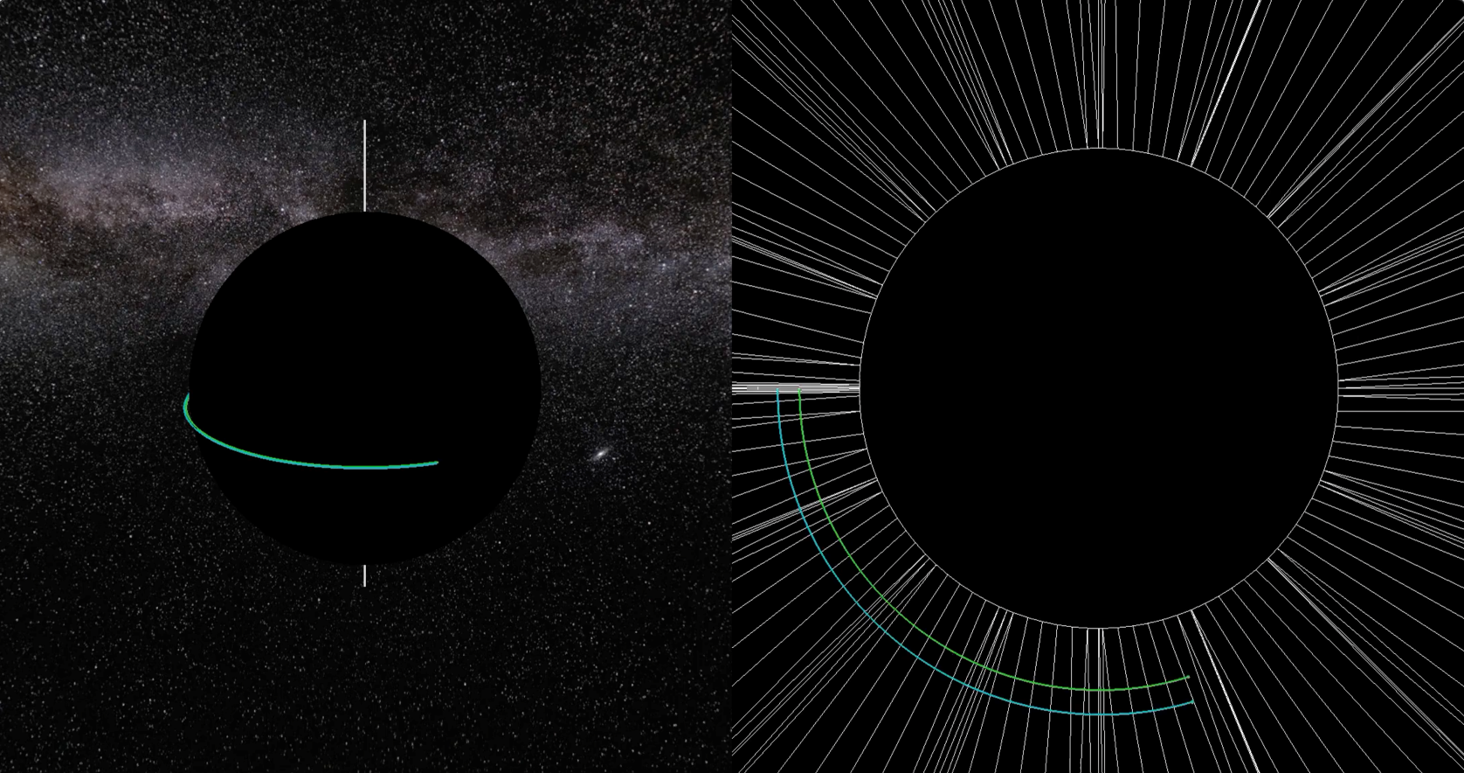}\\
	\includegraphics[width=.75\textwidth]{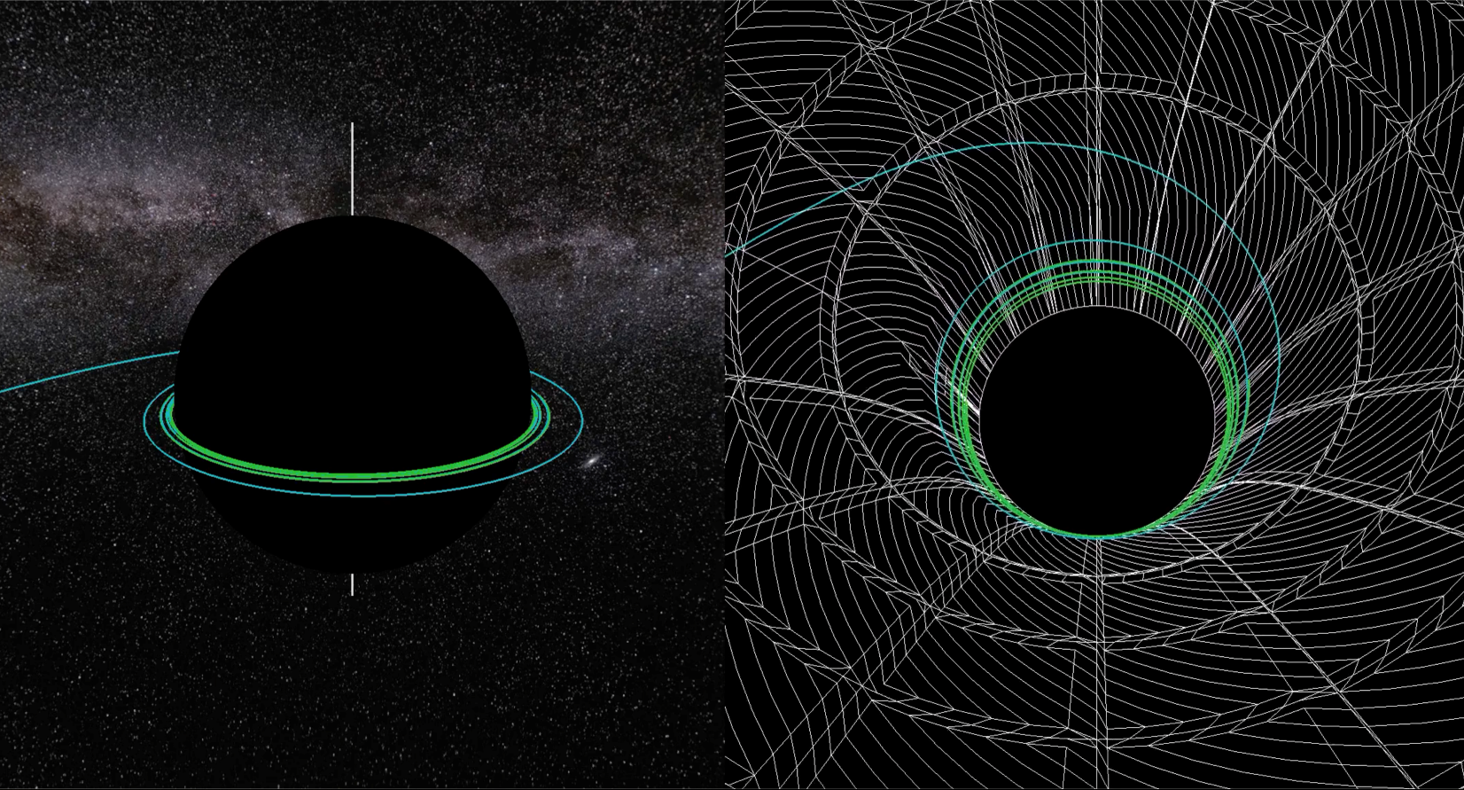}\\
	\includegraphics[width=.75\textwidth]{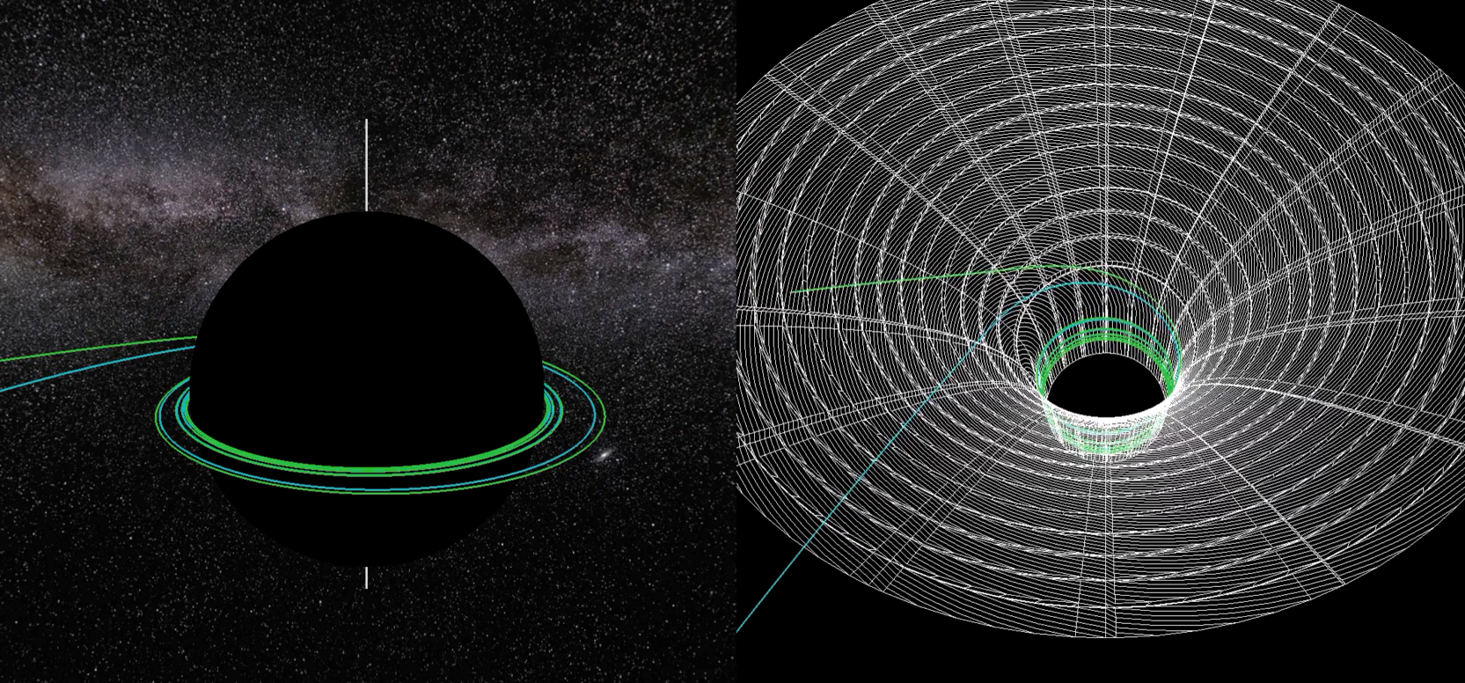}
	\caption{Sequenced frames (top to bottom) from the moving animation of nearby photons, orbiting in a ``real space'' diagram and in an embedding diagram (see App.~\ref{app:Embedding} for its definition).}
	\label{fig:Sequence}
\end{figure}

In the sequence shown in Fig.~\ref{fig:Sequence}, below, two photons orbit a rapidly spinning black hole (with spin $a/M=0.99\%$) together, with the blue one escaping first.
Left panels show the 3D spatial slices in Boyer-Lindquist coordinates; right panels show an embedding diagram of the equatorial plane that illustrates the curvature of the spacetime.

\section{Conclusions}

Visualizing the distribution of the Lyapunov exponent (via the separation of nearby photons) offers us a probe of the chaotic flow of light around a spinning black hole.
We can track the strength of chaoticity through visual simulations, following photons as they course along initially nearby geodesics.
Ultimately, the aim of these visualizations is to articulate the spacetime geometry near the black hole horizon---to offer us intuition into its chaotic behavior.
The Black Hole Explorer could well offer the first empirical glimpse at the Lyapunov exponent of spacetime itself.
It will not be the last.

\acknowledgments      
 
This project/publication is funded in part by the Gordon and Betty Moore Foundation (Grant No. 8273.01).
It was also made possible through the support of a grant from the John Templeton Foundation (Grant No. 62286).
The opinions expressed in this publication are those of the author(s) and do not necessarily reflect the views of these Foundations.
MJ is supported in part by NSF grants AST-2307887.
AL, TG, and RB are supported in part by NSF grants AST-2307888 and PHY-2340457.
BHEX is funded in part by generous support from Mr. Michael Tuteur and Dr. Amy Tuteur, MD.
BHEX is supported by initial funding from Fred Ehrsam.

\clearpage
\appendix

\section{On the making of the embedding diagram}
\label{app:Embedding}

In this appendix, we describe the equations defining the embedding diagram for the equatorial plane of a rotating astrophysical black hole, with mass $M$ and angular momentum $J=aM$.
The spacetime geometry around such a black hole is described by the Kerr metric, which in Boyer-Lindquist coordinates $(t,r,\theta,\phi)$ has the line element
\begin{gather}
	ds^2=-\frac{\Delta}{\Sigma}\left(\ed t-a\sin^2{\theta}\ed\phi\right)^2+\frac{\Sigma}{\Delta}\ed r^2+\Sigma\ed\theta^2+\frac{\sin^2{\theta}}{\Sigma}\left[\left(r^2+a^2\right)\ed\phi-a\ed t\right]^2,\\
	\Delta=r^2-2Mr+a^2,\qquad
	\Sigma=r^2+a^2\cos^2{\theta}.
\end{gather}
This metric is stationary, so the 3D spatial geometry is the same at all times $t$.
As such, the induced geometry on the equatorial plane is obtained by setting $\theta=\pi/2$ and $\ed t=\ed\theta=0$, resulting in a 2D line element
\begin{align}
	\label{eq:Equatorial}
	ds_{\rm eq}^2=\frac{r^2}{\Delta}\ed r^2+\frac{\left(r^2+a^2\right)^2-a^2\Delta}{r^2}\ed\phi^2.
\end{align}
We wish to embed this equatorial geometry into a 3D flat space: the Euclidean geometry $\mathbb{R}^3$, whose metric in cylindrical coordinates $(R,Z,\Phi)$ is
\begin{align}
	\label{eq:Flat}
	ds_{\rm flat}^2=\ed R^2+\ed Z^2+R^2\ed\Phi^2.
\end{align}
Thanks to the azimuthal symmetry of the Kerr metric, we can suppose that the embedding surface $M\subset\mathbb{R}^3$ is given by a 2D surface of revolution parameterized by the coordinates $(r,\phi)$ on the Kerr equatorial plane as
\begin{align}
	R(r,\phi)=R(r),\qquad
	Z(r,\phi)=Z(r),\qquad
	\Phi(r,\phi)=\phi.
\end{align}
Plugging this Ansatz into Eq.~\eqref{eq:Flat}, we obtain an induced metric on the embedding surface $M\subset\mathbb{R}^3$ of the form
\begin{align}
	ds_M^2=\left[\left(\frac{dR}{dr}\right)^2+\left(\frac{dZ}{dr}\right)^2\right]+R^2(r)\ed\phi^2
\end{align}
Setting this equal to Eq.~\eqref{eq:Equatorial} yields the defining equations for the embedding map from the Kerr equatorial plane into the 3D flat Euclidean space $\mathbb{R}^3$:
\begin{align}
	R(r)&=\frac{\sqrt{\left(r^2+a^2\right)^2-a^2\Delta}}{r},
\end{align}
and also
\begin{align}
	Z'(r)=\sqrt{\frac{r^2}{\Delta}-\left(\partial_r\sqrt{\frac{\left(r^2+a^2\right)^2-a^2\Delta}{r^2}}\right)^2}.
\end{align}
For the case of a non-rotating Schwarzschild black hole (with vanishing spin $a=0$), these equations simplify to:
\begin{align}
	R(r)=r,\qquad
	Z(r)=\int\frac{\ed r}{\sqrt{r/(2M)-1}}
	=\sqrt{8M(r-2M)}+Z_0.
\end{align}
In the general case of a spinning Kerr black hole (with $a\neq0$), we obtain $Z(r)$ via numerical integration.

% References
\bibliography{report} % bibliography data in report.bib
\bibliographystyle{spiebib} % makes bibtex use spiebib.bst

\end{document}